\documentclass[12pt,preprint]{aastex}

\begin{document}

\title{Polarization Patterns in Pulsar Radio Emission}
\author{Mark M. McKinnon}
\affil{National Radio Astronomy Observatory,\altaffilmark{1}\altaffiltext{1}
{The National Radio Astronomy Observatory is a facility of the National 
Science Foundation operated under cooperative agreement by Associated 
Universities, Inc.} Socorro, NM \ \ 87801\ \ USA}

\begin{abstract}

A variety of intriguing polarization patterns are created when polarization 
observations of the single pulses from radio pulsars are displayed in a 
two-dimensional projection of the Poincar\'e sphere. In many pulsars, the 
projections produce two clusters of data points that reside at antipodal 
points on the sphere. The clusters are formed by fluctuations in polarization 
amplitude that are parallel to the unit vectors representing the polarization 
states of the wave propagation modes in the pulsar magnetosphere. In other 
pulsars, however, the patterns are more complex, resembling annuli and bow 
ties or bars. The formation of these complex patterns is not understood and 
largely unexplored. An empirical model of pulsar polarization is used to show 
that these patterns arise from polarization fluctuations that are perpendicular 
to the mode vectors. The model also shows that the modulation index of the 
polarization amplitude is an indicator of polarization pattern complexity.
A stochastic version of generalized Faraday rotation can cause the orientation 
of the polarization vectors to fluctuate and is a possible candidate for the 
perpendicular fluctuations incorporated in the model. Alternative models 
indicate that one mode experiences perpendicular fluctuations and the other 
does not, suggesting that the fluctuations could also be due to a mode-selective 
random process, such as scattering in the magnetosphere. A polarization stability 
analysis of the patterns implies that processes intrinsic to the emission are 
more effective in depolarizing the emission than fluctuations in the orientation 
of its polarization vector.

\end{abstract}


\section{INTRODUCTION}

Pulsar radio emission is renowned for its complicated polarization behavior.
The polarization is generally elliptical, yet predominantly linear, and highly 
variable, often switching between two, orthogonally polarized states (e.g. 
Stinebring et al. 1984). A recent trend in the display and interpretation of 
the results from polarization observations of individual pulses has been to 
plot the measured values of the polarization vector's orientation angles at 
a given rotational phase of the pulse in a two-dimensional projection of the 
Poincar\'e sphere. The projections reveal a wide variety of organized 
polarization patterns. For example, towards the center of the pulse in PSR 
B2020+28, the angles reside in a single cluster in one hemisphere of the 
Poincar\'e sphere (McKinnon 2004). The orientation angles form a diffuse 
structure resembling a bow tie near the pulse center of PSR B0818--13 
(Edwards 2004). At the peak of PSR B1133+16, a Hammer-Aitoff projection of 
the angles produces two data clusters, each in a separate hemisphere of the 
Poincar\'e sphere (Karastergiou et al. 2003). The Lambert equal-area 
projections (LEAPs) of the orientation angles in PSR B0329+54 reveal two 
extremes in polarization behavior within the pulsar's pulse (Edwards \& 
Stappers 2004). In the cone emission on the edges of the pulse, the angles 
reside in two, circularly-shaped, bipolar clusters, similar to the angles in 
PSR B1133+16. But within the pulsar's core emission at the pulse center, one 
of the two clusters stretches into an ellipse or bar, while the other spreads 
into an intriguing partial annulus. An accurate description of these patterns 
is needed to determine the physical processes that create them.

The origin of the simple polarization patterns is generally understood 
within the context of the statistical model of McKinnon \& Stinebring (1998, 
2000), aided by the analyses of McKinnon (2004) and Edwards \& Stappers
(2004). A single, compact cluster is produced when the fluctuations in the 
Stokes parameters Q, U, and V are comparable to one another, but less than 
the mean value of the polarization vector's amplitude, as might be expected 
for the measurement of a polarization vector with fixed amplitude and 
orientation accompanied by instrumental noise. Orthogonal polarization modes 
(OPMs) produce the projections showing two clusters of orientation angles. 
The unit vectors representing the mode polarizations are antiparallel to each 
other and form a diagonal in the Poincar\'e sphere. Since the mode vectors are 
antiparallel, the amplitude of the resultant polarization is the difference 
between the mode polarization amplitudes, and the tip of the resultant 
polarization vector will reside in one hemisphere of the Poincar\'e sphere 
or the other depending upon which mode is the stronger of the two. The 
instantaneous orientation of the polarization vector alternates randomly between 
hemispheres due to temporal fluctuations in the polarized intensity of each 
mode that could be caused by the radio emission mechanism or propagation 
effects in the pulsar magnetosphere, such as the birefringence of the modes 
(e.g. Allen \& Melrose 1982; Barnard \& Arons 1986). The resulting polarization 
fluctuations caused by the OPMs are therefore parallel to the mode diagonal.

The clues to the origin of the more complicated polarization patterns are
provided in the analyses of McKinnon (2004) and Edwards \& Stappers (2004), 
the polarization observations of Edwards (2004), and the numerical simulations 
of Melrose et al. (2006). The shape of the Q-U-V data point clusters created 
by the polarization fluctuations is generally an ellipsoid. McKinnon and 
Edwards \& Stappers independently developed a technique to quantify the 
polarization fluctuations by measuring the ellipsoid's dimensions. The 
technique calculates the covariance matrix of the Stokes parameters and
subsequently determines the matrix eigenvalues. The three dimensions of the 
polarization ellipsoid are related to the eigenvalues, and the eigenvectors 
are the three orthogonal axes of the ellipsoid. For some objects, such as 
PSR B0809+74 (Edwards 2004), PSR B1929+10, and PSR B2020+28 (McKinnon 2004), 
two of the eigenvalues are roughly equal but less than the third, which is 
what one would expect for a prolate ellipsoid fashioned by OPMs. But Edwards
(2004) also found examples (e.g. PSR B0320+39 and PSR B0818--13) where all 
three eigenvalues are different, proving that the polarization fluctuations 
possess a component that is {\it perpendicular} to the mode diagonal, in 
addition to the {\it parallel} component produced by the OPMs. Melrose et
al. (2006) had to incorporate fluctuations in the orientation of the mode 
diagonal to replicate the polarization patterns observed by Edwards \& 
Stappers in PSR B0329+54. Randomly varying orientation angles equate to 
both parallel and perpendicular components in the polarization fluctuations. 
Karastergiou et al. (2003) also suggested that random variations in the 
orientation of the mode diagonal might be needed to explain their 
observations of PSR B1133+16. 

McKinnon \& Stinebring (1998; 2000) proposed that the polarization of the 
radio emission is determined by the incoherent superposition of two, highly 
polarized, orthogonal modes. The assumption of highly polarized modes has 
strong theoretical support on the grounds that any plasma has two natural 
modes of wave propagation that are completely polarized (Petrova 2001). The 
assumption of incoherent modes means the modes propagate independently, 
which has two important consequences. First, when the radiation components
are independent, the intensity of the combined radiation is the sum of the 
intensities of the individual radiation components (Chandrasekhar 1960; 
Ishimaru 1978). This realization simplifies the modeling of the radiation 
and its polarization. Second, the independence of the modes requires the 
difference in mode phases to be greater than unity. The mode phase difference, 
$\Delta\chi=\Delta kL$, is the product of the difference between the mode 
wave numbers, $\Delta k$, in the plasma and the distance, $L$, the modes 
propagate through the plasma. Generalized Faraday rotation (GFR) has been
defined as the physical process responsible for creating the difference in 
mode phases (Melrose 1979). If this definition is correct, then GFR must be 
operative in the pulsar magnetosphere. Stochastic GFR can also alter the 
orientation of the polarization vector and, therefore, is a tantalizing 
prospect for the origin of some polarization patterns we observe (Edwards 
\& Stappers 2004).

Melrose et al. (2006) made great strides in simulating the polarization 
patterns with their numerical model and called for further use of 
empirical models to constrain the mechanism that leads to the separation of 
the modes. The objective of this paper is to incorporate the perpendicular 
fluctuations in an analytical model of pulsar polarization in an attempt 
to replicate the observed polarization patterns and identify potential 
processes responsible for the fluctuations. The perpendicular fluctuations 
are incorporated in the model in \S\ref{sec:patterns}. The polarization 
patterns produced by different fluctuation geometries are also determined. 
The fluctuations can depolarize the emission, and, in \S\ref{sec:depol}, the 
depolarization is quantified with a polarization stability factor for most 
of the fluctuation geometries. In \S\ref{sec:perp}, the perpendicular 
fluctuations are interpreted in the context of stochastic GFR or scattering 
in the pulsar magnetosphere. The results and implications of the analysis 
are discussed in \S\ref{sec:discuss}. Conclusions are summarized in 
\S\ref{sec:conclude}. The Appendix lists the joint probability densities 
of the polarization vector's orientation angles for each of the fluctuation 
geometries. 

\section{ANALYTICAL MODEL OF POLARIZATION PATTERNS}
\label{sec:patterns}

The polarization patterns are two-dimensional representations of the joint 
probability density of the polarization vector's longitude, $\phi$, and 
colatitude, $\theta$. The functional form of the joint probability density 
can be estimated from a statistical model of pulsar polarization. An 
analytical, empirical model of pulsar polarization has been summarized in 
McKinnon (2003). Here, the model is generalized to accommodate polarization 
fluctuations perpendicular to the mode diagonal.

The Stokes parameters Q, U, and V completely describe the polarization of 
the radiation. The measured values of Q, U, and V are the linear sums of 
the pulsar-intrinsic polarization fluctuations and the instrumental noise. 
An analytical description of many types of polarization patterns can be 
determined by assuming that all fluctuations are independent, normal random 
variables (RVs). Since the sum of independent, normal RVs is also a normal 
RV, this assumption allows us to interpret the polarization as being 
composed of fixed and fluctuating parts, where the fluctuations follow a 
normal distribution. The fixed part is the mean value, $\mu$,  of the 
polarization vector amplitude. The analysis can be simplied by defining a 
new Cartesian coordinate system, q, u, and v, within the Poincar\'e sphere 
where the new v-axis is aligned with the mean orientation of the polarization 
vector. The simple equations that specify the model are then
\begin{equation}
{\rm q} = x_{\rm q},
\label{eqn:q}
\end{equation}
\begin{equation}
{\rm u} = x_{\rm u},
\label{eqn:u}
\end{equation}
\begin{equation}
{\rm v} = \mu + x_{\rm v}.
\label{eqn:v}
\end{equation}
The normal RVs $x$ account for the polarization fluctuations in each of q, u, 
and v. They each have a zero mean, and their standard deviations are denoted 
by $\sigma_q, \sigma_u$, and $\sigma_v$ in what follows. The q, u, v coordinate
system forms the eigenbasis of the polarization ellipsoid, and $\sigma_q, 
\sigma_u$, and $\sigma_v$ are the square roots of its eigenvalues (McKinnon 
2004; Edwards \& Stappers 2004). Equations~\ref{eqn:q}-\ref{eqn:v} give the 
Cartesian coordinates that define the instantaneous orientation and amplitude 
of the polarization vector. The joint probability density of the vector's 
orientation angles can be computed by a transformation from Cartesian to 
spherical coordinates. 

Five cases with different fluctuation geometries, from simple to complex, are 
explored in the subsections that follow. Four cases investigate possible 
geometries where at least two of the eigenvalues are equal. The fifth case is 
more general, and evaluates a scenario where all the eigenvalues are different.
Granted, other cases could be considered, but they generally produce patterns 
that are simple rotations of the patterns produced here. The cases show that 
the polarization pattern is determined by the relative magnitudes
of $\sigma_q, \sigma_u$, and $\sigma_v$ and the value of $\mu$. 

Despite the simplicity of Equations~\ref{eqn:q}-\ref{eqn:v}, the joint density 
produced from them has a complicated mathematical form. Interestingly, but 
quite understandably, the mathematical form of the joint density is basically 
the same in all cases, differing only in a parameterization determined by the 
geometry of the polarization fluctuations. The form of the joint density and 
its geometrical parameterization for all cases considered is given in the 
Appendix. The functional form of the joint density can be captured in a much 
simpler conditional density, which is the probability density of the orientation 
angles at a fixed value of the polarization amplitude, $r_o$. The detailed 
procedure for calculating both the joint and conditional densities is 
described in McKinnon (2003, 2006) and is not reproduced here. The conditional 
probability densities of the orientation angles are derived for each case 
below. 

Table 1 summarizes the results obtained for each case by listing the 
resulting shape of the q-u-v ellipsoid and the formal name of the orientation 
angles' conditional density. The axis orientation in the table describes the 
orientation of the ellipsoid's axis of symmetry with respect to the v axis. 

\begin{figure}
\plotone{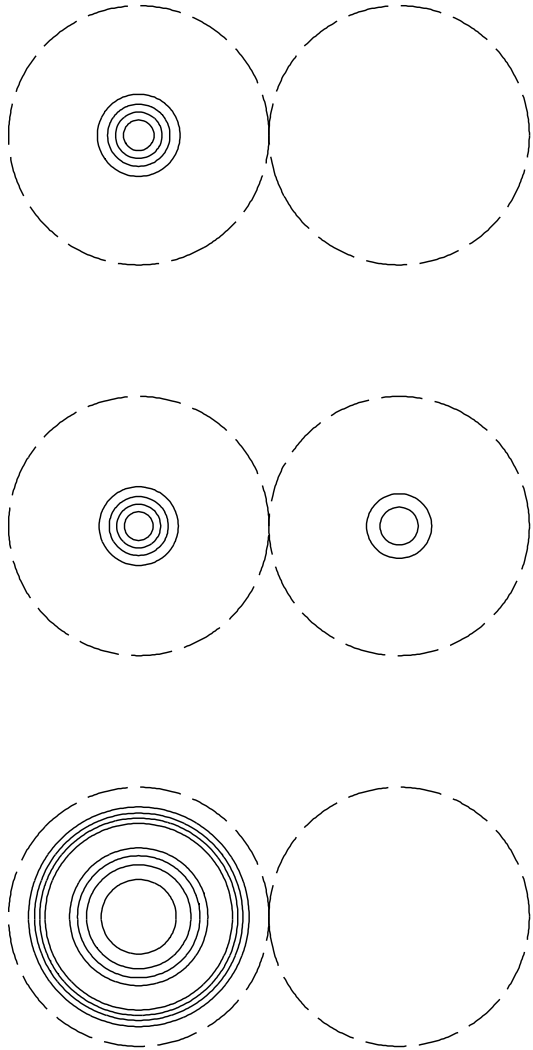}
\caption{Examples of possible polarization patterns in pulsar radio emission.
The patterns are shown as contour plots of Lambert equal area projections of 
the Poincar\'e sphere. {\it Top panel:} Fisher distribution with $\kappa = 4$ 
(Case 1). {\it Middle panel:} Bingham-Mardia bipolar distribution with 
$\kappa = 3, \gamma = 0.015$ created by fluctuations parallel to the 
polarization vector (Case 2). {\it Bottom panel:} Bingham-Mardia girdle 
distribution with $\kappa=4,\gamma=0.6$ created by fluctuations perpendicular 
to the polarization vector (Case 3). Contour levels are at 0.2, 0.4, 0.6,
and 0.8 of the peak value in each projection.}
\label{fig:pattern1}
\end{figure}

Figures~\ref{fig:pattern1} and~\ref{fig:pattern2} show LEAP examples for each 
case. A LEAP is simply a polar plot where a data point's azimuth is $\phi$ 
and its radius is $2\sin(\theta/2)$. The most attractive feature of a LEAP 
is it preserves the density of data points on the sphere when projecting 
them in two dimensions (Fisher et al. 1987). In the figures, the left side 
of the LEAP is the projection of the top hemisphere of the q-u-v sphere as 
viewed down the v axis, where $\theta=0$. The right side of the LEAP is the 
projection of the bottom hemisphere at $\theta=\pi$. The circular edge of 
the projections is the equator of the q-u-v sphere, where $\theta=\pi/2$. 

\subsection{Case 1: $\sigma_q = \sigma_u = \sigma_v$}

The first case to consider is the simplest of the five, and evaluates the 
conditional density when the standard deviations of the polarization 
fluctuations are identical (i.e. $\sigma_q = \sigma_u = \sigma_v = \sigma$). 
The shape of the q-u-v cluster in this case is a spheroid because the standard 
deviations are equal. Since the cluster is a spheroid, it does not have a
unique symmetry axis. This case was evaluated by McKinnon (2003), and 
represents the statistics of a fixed polarization vector accompanied by 
instrumental noise. The conditional density of the polarization vector's 
orientation angles is a Fisher distribution
\begin{equation}
f_1(\theta,\phi |r_o) = {\sin\theta\over{4\pi}}
                      {\kappa^2\exp(\kappa^2\cos\theta)\over{\sinh(\kappa^2)}},
\label{eqn:cond1}
\end{equation}
where 
\begin{equation}
\kappa^2={\mu r_o\over{\sigma^2}}.
\label{eqn:kappa1}
\end{equation} 
The parameter $\kappa$\footnote{The equivalent of Equation~\ref{eqn:kappa1} 
in McKinnon (2003) is used to define $\kappa$, instead of $\kappa^2$. The 
definition used here is preferred because it implies a connection with 
variance.} is inversely related to the conditional density's standard 
deviation. Since $\sin\theta$ is proportional to the derivative of 
$\cos\theta$, the Fisher distribution is an exponential in $\cos\theta$. 
The density is not a function of $\phi$ because the distribution of data 
points is azimuthally symmetric about the mean orientation of the polarization 
vector (i.e. $\phi$ is uniformly distributed over $2\pi$ and is independent 
of $\theta$). Therefore, the contour shapes in a LEAP for this case are 
always circles. 

The conditional density in Equation~\ref{eqn:cond1} is a probability density 
function (PDF), but the equation to plot in a LEAP is the probability density 
element (PDE). For spherical data, the PDE is the PDF without the leading
$\sin\theta$ term (Fisher et al. 1987; see also Edwards \& Stappers 2004). 
When $\kappa>1$, the polarization pattern derived from the PDE is a single 
set of concentric, circular contours with a peak at the center of the left 
LEAP hemisphere, where $\theta = 0$ (top panel of Figure~\ref{fig:pattern1}). 
When $\kappa<1$, the conditional density becomes isotropic, and the LEAP 
will show data points uniformly scattered over both hemispheres. The joint 
probability density of the orientation angles (Equation~\ref{eqn:joint}) is 
parameterized solely by the polarization signal-to-noise ratio, $s=\mu/\sigma$. 
The functional form of the joint density is similar to that of the conditional 
density. 

An observational example of this case can be found towards the center of the 
pulse in PSR B2020+28, as shown in the top panel of Figure~\ref{fig:PSRpattern}.
The polarization pattern is a set of circularly-shaped contours in a single 
hemisphere of the LEAP. The cluster of polarization data points at this pulse 
location is clearly a spheroid because its three dimensions are nearly identical 
(see Figures 3 and 5 of McKinnon 2004). 

\subsection{Case 2: $\sigma_q = \sigma_u < \sigma_v$}

The second case investigates fluctuations in q and u that are equal, 
$\sigma_q = \sigma_u =\sigma$, but less than those in v by a factor of 
$(1+\rho^2)^{1/2}$, where $\rho\ge 0$. This case was evaluated in McKinnon 
(2006), and considers a system dominated by fluctuations along the 
polarization vector, as is caused by OPMs. The shape of the q-u-v data point 
cluster created by these polarization fluctuations is a prolate ellipsoid. 
Its major axis is parallel to v. The conditional density of the vector's 
orientation angles is a Bingham-Mardia (BM) bipolar distribution
\begin{equation}
f_2(\theta,\phi |r_o) = {\sin\theta\over{2\pi}}
              {\exp[\kappa^2(\cos\theta + \gamma)^2]\over{w_2(\kappa,\gamma)}}.
\label{eqn:cond2}
\end{equation}
The distribution is parameterized by the constants $\kappa$ and $\gamma$.
\begin{equation} 
\kappa^2 = {r_o^2\rho^2\over{2\sigma^2(1+\rho^2)}} 
\label{eqn:kappa2}
\end{equation} 
\begin{equation}
\gamma = {\mu\over{r_o\rho^2}}
\label{eqn:gamma2}
\end{equation}
The constant $w_2(\kappa,\gamma)$ normalizes the density and is found by 
integrating the numerator of Equation~\ref{eqn:cond2}. The conditional 
density becomes the Watson bipolar distribution when $\mu=0$ (McKinnon 2006). 

The conditional density can be unimodal or bimodal depending upon the value 
of $\gamma$. When $\gamma > 0$, the PDE always peaks at the center of the 
left LEAP hemisphere where $\theta=0$. When $|\gamma|\ll 1$ the polarization 
pattern is bimodal (i.e. a secondary peak appears at the center of the right 
LEAP hemisphere where $\theta=\pi$). The patterns are unimodal for larger 
values of $\gamma$. As with Case 1, the shapes of the density's contours are 
always circles because of the azimuthal symmetry of the problem. The middle 
panel of Figure~\ref{fig:pattern1} shows an example of the PDE from 
Equation~\ref{eqn:cond2}. 

As shown in McKinnon (2006), the functional form of the joint probability 
density (Equation~\ref{eqn:joint}) is very similar to that of the conditional 
density. The polarization modulation index, $\beta$, determines whether the 
joint density is bimodal or unimodal. In this case, the modulation index is 
defined by
\begin{equation}
\beta = {(\sigma_v^2-\sigma^2)^{1/2}\over{\mu}} = {\rho\over{s}},
\end{equation}
where $s=\mu/\sigma$ is the ratio of the polarization amplitude to the 
effective noise of the system. The fluctuations along v that are over 
and above the effective noise are represented by $\rho\sigma$. When 
$\beta > 1$, the fluctuations along v exceed the mean polarization, and 
the joint density is bimodal with data points residing in both LEAP 
hemispheres.  When $\beta < 1$, the mean exceeds the fluctuations, and the 
joint density is unimodal.

An observational example of this case occurs on the trailing edge of the 
pulse of PSR B1929+10 (middle panel of Figure~\ref{fig:PSRpattern}). There,
the polarization pattern consists of a set of circularly-shaped contours in 
each hemisphere of the LEAP. The contours in the right hemisphere are not 
centered in the projection because the modes are not precisely orthogonal 
(McKinnon 2004). The cluster of polarization data points at this location 
in the pulse has the form of a prolate ellipsoid because two of its dimensions 
are equal but smaller than the third (Figures 4 and 5 of McKinnon 2004). Another 
example of this case occurs in the cone emission of PSR B0329+54. Again, the 
pattern consists of a set of circularly-shaped contours in each hemisphere of
the LEAP (Figure 2 of Edwards \& Stappers 2004), and the relative dimensions 
of the data point clusters are consistent with those expected for a prolate 
ellipsoid (Figure 3 of Edwards \& Stappers).

\begin{figure}
\plotone{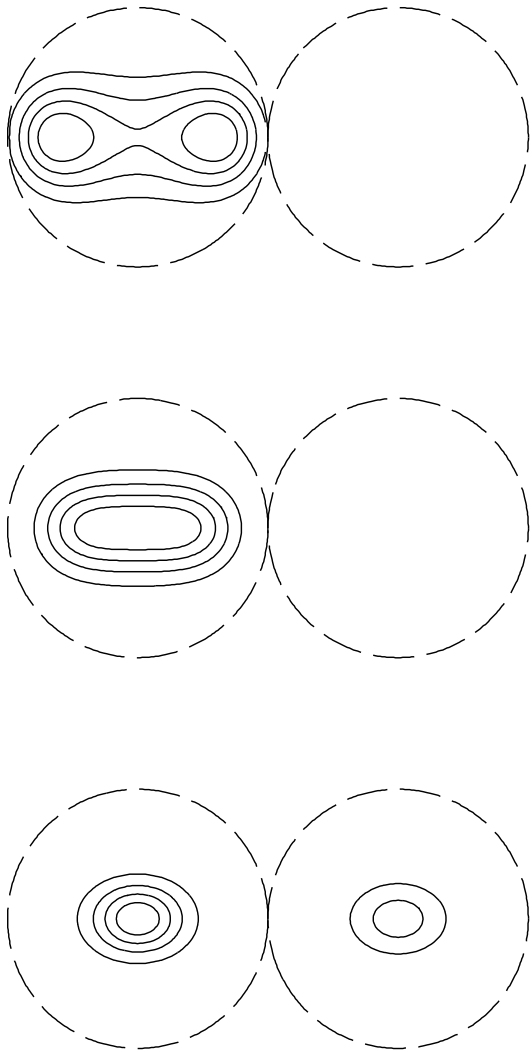}
\caption{Additional examples of possible polarization patterns in pulsar radio 
emission. {\it Top panel:} A hybrid of the Bingham-Mardia distribution where 
the polarization fluctuations are primarily along q (Case 4). This particular 
pattern was formed with $\kappa = 2$ and $\gamma=0.65$. {\it Middle panel:} 
Same as the top panel but with $\gamma = 1$. {\it Bottom panel:} Another 
hybrid of the Bingham-Mardia distribution where the fluctuations in all three 
dimensions are different, with $\kappa_o = -4, \kappa = 3,$ and 
$\gamma=-0.04$ (Case 5). Contour levels are at 0.2, 0.4, 0.6, and 0.8 of the 
peak value in each projection.}
\label{fig:pattern2}
\end{figure}

\subsection{Case 3: $\sigma_q = \sigma_u > \sigma_v$}

To model fluctuations perpendicular to the polarization vector, the 
fluctuations in q and u are taken to be equal, 
$\sigma_q = \sigma_u =\sigma(1+\eta^2)^{1/2}$, but greater than those in 
v, where $\sigma_v =\sigma$. The constant $\eta$ denotes the magnitude 
of the fluctuations perpendicular to the polarization vector, while $\rho$ 
is reserved to signify the parallel fluctuations. The polarization fluctuations 
in this case create a q-u-v data point cluster in the shape of an oblate 
ellipsoid. The minor axis of the ellipsoid is parallel to v. This case does 
not preclude the possibility of OPMs; it simply stipulates that fluctuations
perpendicular to the polarization vector exceed the parallel fluctuations. 
The conditional density is a BM girdle distribution
\begin{equation}
f_3(\theta,\phi |r_o) = {\sin\theta\over{2\pi}}
         {\exp[-\kappa^2(\cos\theta - \gamma)^2]\over{w_3(\kappa,\gamma)}},
\label{eqn:cond3}
\end{equation}
where
\begin{equation}
\gamma = {\mu(1+\eta^2)\over{r_o\eta^2}}
\label{eqn:gamma3}
\end{equation}
and $\kappa$ has the same definition as in Case 2, Equation~\ref{eqn:kappa2}, 
but with $\rho$ replaced by $\eta$. The constant $w_3(\kappa,\gamma)$ is
a normalization factor. The conditional density is normal in $\cos\theta$ with 
mean $\gamma$ and a standard deviation proportional to $1/\kappa$. Since the 
density is a distribution of $\cos\theta$, which must lie in the range 
$0\le|\cos\theta|\le 1$, a strict mathematical interpretation of 
Equation~\ref{eqn:cond3} requires $\gamma$ to lie in the range 
$0\le|\gamma|\le 1$. A variant of Equation~\ref{eqn:cond3} has been used to 
model the motion of volcanic hotspots on the surface of the Earth (Bingham 
\& Mardia 1978; Mardia \& Gadsden 1977). 

As shown in the bottom panel of Figure~\ref{fig:pattern1}, the polarization 
pattern formed from the conditional density PDE is an annulus in a single 
LEAP hemisphere. The PDE peaks at $\cos\theta=\gamma$ for all longitudes, 
again because of the azimuthal symmetry of the problem. The outer edge of the 
annulus has a steep slope and its inner edge has a more gradual slope. The 
sign of $\gamma$ determines the hemisphere where the annulus resides. When 
$\gamma=0$, the PDE peaks at $\theta=\pi/2$, and the pattern contours are 
concentrated around the equator in each LEAP hemisphere. As $\gamma$ approaches 
unity, the annulus collapses into a single cluster with a peak at $\theta=0$. 
Unlike the density in Case 2 that can be bimodal, the density in this case 
is always unimodal. 

The shape of the polarization pattern derived from the joint probability 
density (Equation~\ref{eqn:joint}) is again determined by the polarization 
modulation index, here defined as $\beta=\eta/s$. The polarization pattern 
is an annulus when the polarization fluctuations are comparable to the mean 
polarization ($\beta\simeq 1$). The pattern is a single peak at $\theta=0$
when the polarization fluctuations are less than the mean ($\beta<1$).

The polarization annulus was first recognized by Melrose et al. (2006) in 
their attempt to model the polarization pattern observed by Edwards \& Stappers
(2004) in the core emission of PSR B0329+54. Their numerical model showed 
that the annulus formed when the polarization amplitude was weak. The 
analytical model developed here reproduces a similar pattern precisely when 
$\mu$ (i.e. $\gamma$) is small or when $\beta\simeq 1$.

\subsection{Case 4: $\sigma_q > \sigma_u = \sigma_v$}

The fluctuation geometries described in the preceding cases have all been 
symmetric in azimuth. The symmetry is broken in this case by setting the 
fluctuations in u and v equal to one another, $\sigma_v = \sigma_u =\sigma$, 
but less than those in q, $\sigma_q = \sigma(1+\eta^2)^{1/2}$. Now the shape 
of the q-u-v data point cluster is a prolate ellipsoid with its major axis 
oriented along q and thus perpendicular to v. The conditional density is a 
hybrid of the BM distribution
\begin{equation}
f_4(\theta,\phi |r_o) = {\sin\theta\over{2\pi}}
           {\exp[-\kappa^2(\cos\theta  - \gamma)^2]
           \exp(-\kappa^2\sin^2\phi\sin^2\theta)\over{w_4(\kappa,\gamma)}},
\label{eqn:cond4}
\end{equation}
where $\kappa$ and $\gamma$ have the same definitions as in Case 3, and
$w_4(\kappa,\gamma)$ is a constant that normalizes the distribution. 
Equation~\ref{eqn:cond4} is the conditional density of Case 3 (first 
exponential term in the equation) that is further shaped by a 
longitude-dependent term (the second exponential term). The PDE resides in 
one hemisphere of the LEAP and generally peaks at two locations where 
$\cos\theta=\gamma$ and $\phi = 0,\pi$. As shown in the top and middle 
panels of Figure~\ref{fig:pattern2}, the pattern has a bar shape when 
$\gamma=1$, and a bow tie shape when $\gamma<1$. For even smaller values 
of $\gamma$, the bow tie separates into two distinct peaks.

Again, the polarization modulation index, $\beta=\eta/s$, determines the 
pattern shape derived from the joint probability density. When $\beta>1$,
the pattern resembles a bow tie. The pattern is a bar when $\beta<1$.

\begin{figure}
\plotone{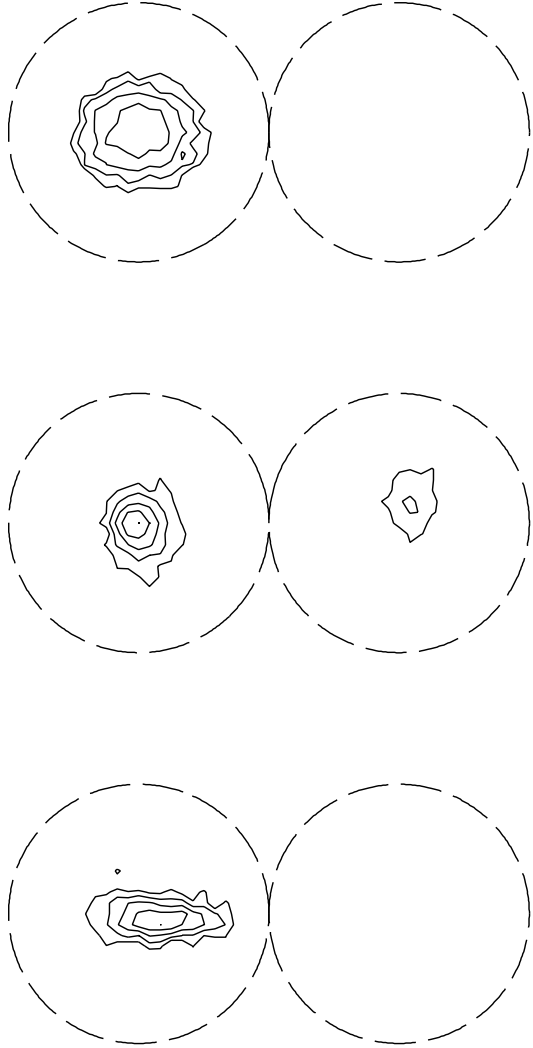}
\caption{Lambert equal area projections of the observed orientation angles 
of the polarization vector in three pulsars. {\it Top panel:} Polarization 
pattern measured towards the center of PSR B2020+28. Contour levels are -12, 
-9, -6, and -3 decibels (dB) referenced to 0 dB at the peak of the projection. 
{\it Middle panel:} Pattern measured on the trailing edge of PSR B1929+10.
Contours are -20, -15, -10, and -5 dB. {\it Bottom panel:} Pattern measured 
near the peak of PSR B0823+26. Contours are -8, -6, -4, and -2 dB. The data 
were recorded by Stinebring et al. (1984) with the Arecibo radio telescope at
1404 MHz.}
\label{fig:PSRpattern}
\end{figure}

An observational example of this case appears near the peak of PSR B0823+26
(bottom panel of Figure~\ref{fig:PSRpattern}). The polarization pattern 
consists of a set of highly elongated contours in a single hemisphere of the
LEAP. Another observational example of this case may occur in the precursor 
to the core component of PSR B0329+54, as shown in Figure 2 of Edwards \& 
Stappers (2004). There, an ellipitical bar appears superimposed upon a noise 
background in the left LEAP hemisphere. Only the noise background of uniformly 
distributed data points appears in the right hemisphere. As shown in their 
Figure 3, two of the dimensions of the Q-U-V cluster are equal but smaller 
than the third; therefore, the shape of the cluster is a prolate ellipsoid. 
For the polarization pattern to be confined primarily to one LEAP hemisphere, 
the ellipsoid's major axis must be perpendicular to the mode diagonal, as 
required in this case.

\subsection{Case 5: $\sigma_u < \sigma_q = \sigma_v$}

The final and most complex, yet general, geometry to consider is when 
the fluctuations in q and v are greater than those in u, but not 
necessarily equal to one another. Here, the fluctuations in u are defined
as $\sigma_u =\sigma$, and the fluctuations in q and v are
$\sigma_q = \sigma(1+\eta^2)^{1/2}$ and $\sigma_v = \sigma(1+\rho^2)^{1/2}$,
respectively. Generally, the shape of the q-u-v ellipsoid in this case is 
an irregular ellipsoid that does not have an axis of symmetry. When 
$\eta=\rho$, the ellipsoid is oblate with its minor axis perpendicular to v,
and the conditional density is
\begin{equation}
f_5(\theta,\phi |r_o) = {\sin\theta\over{2\pi}}
           {\exp[\kappa^2(\cos\theta + \gamma)^2]
           \exp(\kappa^2\sin^2\phi\sin^2\theta)
           \over{w_5(\kappa,\gamma)}}
\end{equation}
where $\kappa$ and $\gamma$ are now defined by Equations~\ref{eqn:kappa2} 
and~\ref{eqn:gamma2}, respectively, under Case 2, and $w_5(\kappa,\gamma)$ 
is a normalization factor. 

When $\eta\neq\rho$, the conditional density is 
\begin{equation}
f_5(\theta,\phi |r_o) = {\sin\theta\over{2\pi}}
           {\exp[-\kappa_o(\cos\theta - \gamma)^2]
           \exp(-\kappa^2\sin^2\phi\sin^2\theta)
           \over{w_5(\kappa,\kappa_o,\gamma)}}
\label{eqn:cond5}
\end{equation}
where $\kappa$ has the same definition as in Cases 3 and 4, and $\kappa_o$ 
and $\gamma$ are given by

\begin{equation}
\kappa_o = {r_o^2(\eta^2-\rho^2)\over{2\sigma^2(1+\rho^2)(1+\eta^2)}}
\end{equation}

\begin{equation}
\gamma = {\mu(1+\eta^2)\over{r_o(\eta^2-\rho^2)}}
\label{eqn:gamma5}
\end{equation}
Notice that $\kappa_o$ and $\gamma$ can be negative depending upon the 
values of $\eta$ and $\rho$. Equation~\ref{eqn:cond5} is very general in that 
it reproduces the conditional density for Case 4 (Equation~\ref{eqn:cond4}) 
when $\rho=0$. Similarly, it becomes the conditional density for Case 2 
(Equation~\ref{eqn:cond2}) when $\eta=0$. When $\rho\ge\eta$ such that 
$|\gamma|<1$, the polarization pattern derived from Equation~\ref{eqn:cond5} 
is bimodal and very similar to that of Case 2, with the exception that the 
PDE contours now have elliptical, instead of circular, cross sections because 
of the asymmetry introduced by the fluctuations in q (see the bottom panel 
of Figure~\ref{fig:pattern2}). When $\eta > \rho$, the polarization pattern 
can be a bar or bow tie in a single hemisphere, as in Case 4, depending upon 
the value of $\gamma$.

An observational example of this case can be found in PSR B0818--13, where
the LEAP at the pulse peak is a diffuse structure that resembles a bow tie 
(Figure 5 of Edwards 2004). The cluster dimensions are not equal to one 
another, but two of the dimensions are noticeably larger than the third 
(Edwards, Figure 3), suggesting that the cluster shape is an irregular 
ellipsoid, as this case requires. To get the bow tie shape in its LEAP, the 
fluctuations along q would have to be slightly larger than those along v.

\section{DEPOLARIZATION BY THE FLUCTUATIONS}
\label{sec:depol}

The polarization patterns observed in pulsars are indicators of how the 
orientation angles of a polarization vector fluctuate on the Poincar\'e 
sphere. The fluctuations are polarization instabilities that can depolarize 
the emission. The conditional densities derived in \S\ref{sec:patterns} 
replicate many of the observed polarization patterns, and can be used to 
quantify the degree of depolarization caused by the fluctuations.
 
Manchester et al. (1973) first observed that the percentage linear 
polarization of pulsar radio emission decreased with increasing radio 
frequency, and Manchester et al. (1975) suggested that the depolarization may 
arise from an increase in the randomization of polarization position angle. 
Manchester et al. (1975) and Cordes \& Hankins (1977) quantified the stability 
of the linear polarization with a two-dimensional polarization stability 
factor, here defined as
\begin{equation}
\sigma_p = \Biggl({\langle Q\rangle^2 +\langle U\rangle^2\over
      {\langle Q^2\rangle + \langle U^2\rangle}}\Biggr)^{1/2}.
\end{equation}
Assuming that the position angle, $\psi$, is a normal RV with a zero mean 
and a standard deviation, $\sigma_\psi$, Cordes \& Hankins (1977) calculated 
the moments of Q and U to derive a polarization stability factor given by
\begin{equation}
\sigma_p=\exp(-\sigma_\psi^2/2).
\end{equation}
An identical expression is used to describe the depolarization caused by
stochastic Faraday rotation in the interstellar medium (e.g. Spangler 1982;
Melrose \& Macquart 1998). 

\begin{figure}
\plotone{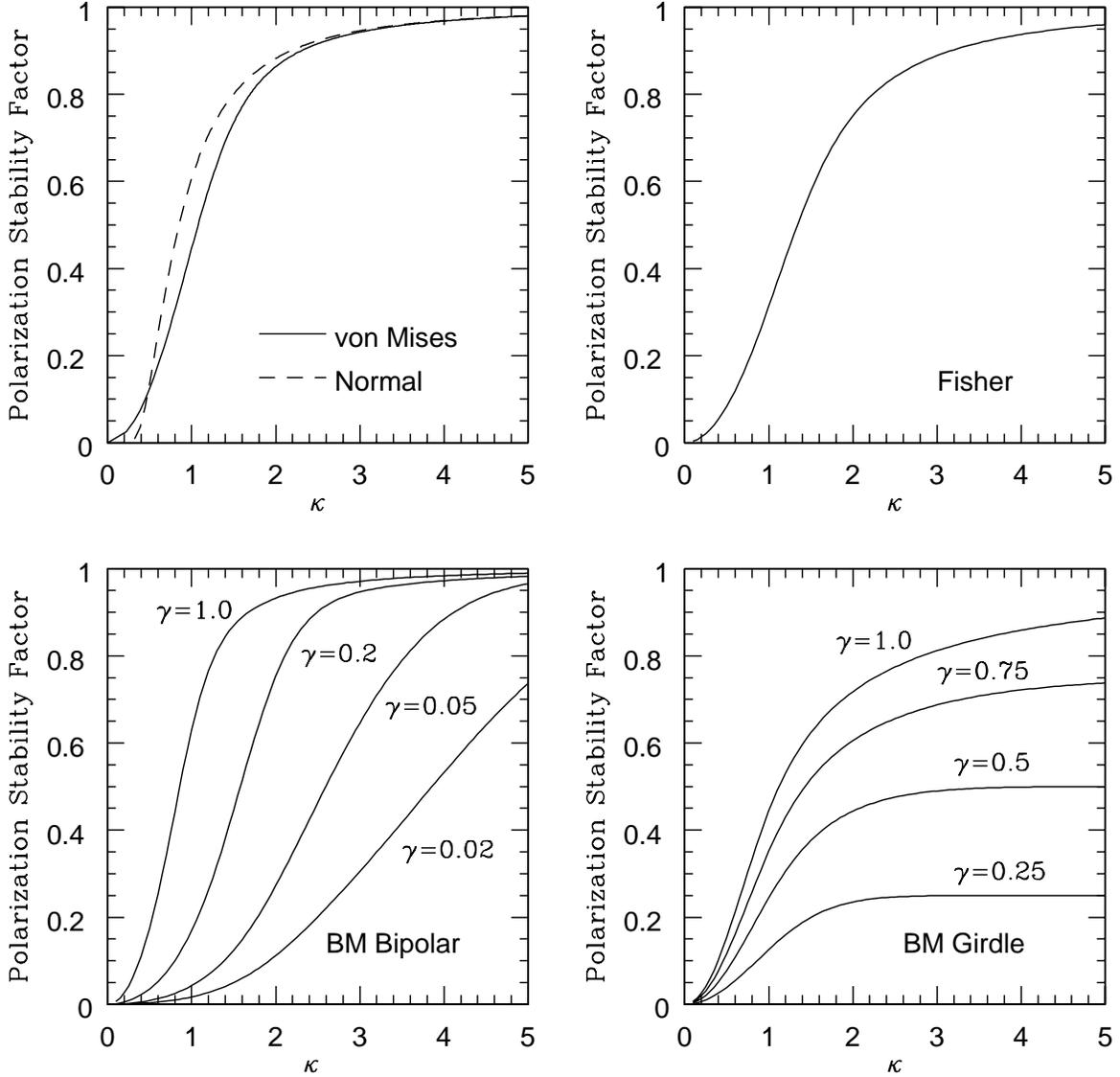}
\caption{The dependence of the polarization stability factor upon
different types of fluctuations in the position angle or colatitude of
the polarization vector. The top left panel shows the stability factor 
when the fluctuations in position angle follow the two-dimensional von 
Mises and normal distributions. The top right panel shows the stability 
factor when the colatitude of the polarization vector fluctuates 
according to the three-dimensional Fisher distribution (Case 1). The 
stability factors determined from the Bingham-Mardia bipolar distribution 
(Case 2) and the Bingham-Mardia girdle distribution (Case 3) are shown 
in the bottom left and right panels, respectively, for different values 
of $\gamma$. The factors are plotted versus $\kappa$, which is related 
inversely to the dispersion in colatitude or position angle.}
\label{fig:psf}
\end{figure}

Using the same methodology described in \S\ref{sec:patterns}, McKinnon (2003) 
showed that the conditional density of the position angle follows a von Mises 
distribution
\begin{equation}
f(\psi|r_o)={\exp(\kappa^2\cos2\psi)\over
                   {\pi\rm{I_0}(\kappa^2)}},
\end{equation}
where $\rm{I_0}(x)$ is the modified Bessel function of order zero and 
$\kappa$ is inversely related to the position angle dispersion. The 
polarization stability factor derived from the von Mises distribution is 
\begin{equation}
\sigma_p = {\rm{I_1}(\kappa^2)\over{\rm{I_0}(\kappa^2)}},
\end{equation}
where $\rm{I_1}(x)$ is the modified Bessel function of first order. When 
the fluctuations in $\psi$ are small ($\kappa\gg 1$), the von Mises 
distribution is almost indistinguishable from a normal distribution 
with a standard deviation of $\sigma_\psi = (2\kappa)^{-1}$. Of the two 
distributions, the von Mises distribution is the more accurate representation 
of position angle fluctuations because $\psi$ is distributed on a semi-circle, 
not a line, and lies in the range $0\le\psi<\pi$, instead of 
$-\infty\le \psi < \infty$ (Fisher et al. 1987). The stability factors 
derived from the two distributions are compared in the top left panel of 
Figure~\ref{fig:psf}.

The observations of Edwards (2004) and Edwards \& Stappers (2004), however,
have clearly shown that the fluctuations occur in all three Stokes parameters,
and not just in Q and U. Consequently, polarization fluctuations in the 
three dimensional case cause the vector colatitude to vary in addition to 
its longitude (position angle). The depolarization can be quantified by 
extending the definition of the polarization stability factor to 
\begin{equation}
\sigma_p=\Biggl({\langle Q\rangle^2 + \langle U\rangle^2 + \langle V\rangle^2
        \over{\langle Q^2\rangle + \langle U^2\rangle + \langle V^2\rangle}}
        \Biggr)^{1/2}.
\end{equation}
For any joint probability density of $\theta$ and $\phi$ that is properly
normalized, the sum of the second moments of the Stokes parameters is 
constant because of the definitions of the Stokes parameters and 
trigonometric identities. The first moments of Q and U are equal to zero 
for all the distributions derived in \S\ref{sec:patterns}. Therefore, the
only term that contributes to the stability factor is the first moment of
V.

The polarization stability factor derived from the Fisher distribution 
(Case 1) is
\begin{equation}
\sigma_p = \coth(\kappa^2)-{1\over{\kappa^2}},
\end{equation}
and is shown in the top right panel of Figure~\ref{fig:psf}. Very little
depolarization occurs when the fluctuations are small ($\kappa\gg 1$),
but the depolarization can be significant when the fluctuations are large
($\kappa\simeq 1$).

The polarization stability factor derived from the BM bipolar distribution 
(Case 2) is 
\begin{equation}
\sigma_p = {\exp[\kappa^2(1+\gamma^2)]\sinh(2\kappa^2\gamma)
           \over{\kappa^2 w_2(\kappa,\gamma)}} - \gamma,
\end{equation}
where the constant $w_2(\kappa,\gamma)$ is given by
\begin{equation}
w_2(\kappa,\gamma)=\int_{\gamma-1}^{\gamma+1}\exp(\kappa^2x^2)dx.
\end{equation}
The stability factor for Case 2 is plotted against $\kappa$ for different 
values of $\gamma$ in the bottom left panel of Figure~\ref{fig:psf}. The 
fluctuations can depolarize the radiation, but the figure shows that the 
depolarization is more pronounced for small values of $\gamma$.

The polarization stability factor derived from the BM girdle distribution 
(Case 3) is 
\begin{equation}
\sigma_p = \gamma - {\exp[-\kappa^2(1+\gamma^2)]\sinh(2\kappa^2\gamma)
           \over{\kappa^2 w_3(\kappa,\gamma)}}.
\end{equation}
The constant $w_3(\kappa,\gamma)$ is
\begin{equation}
w_3(\kappa,\gamma) = {\sqrt{\pi}\over{2\kappa}}\{{\rm erf}[\kappa(1-\gamma)]
                   + {\rm erf}[\kappa(1+\gamma)]\}.
\end{equation}
The stability factor for Case 3 is shown in the bottom right panel of 
Figure~\ref{fig:psf}. As with Case 2, the depolarization is more severe
for small values of $\gamma$. The stability factor approaches 
$\sigma_p = \gamma$ when $\kappa\gg 1$.

The effect of $\kappa$ and $\gamma$ on the frequency-dependent polarization 
can be summarized as follows. The term $\kappa$ represents the magnitude of 
the fluctuations and sets the overall size of the polarization pattern; 
larger patterns are formed from smaller values of $\kappa$. The frequency 
dependence of $\kappa$ may be set by the process responsible for the 
polarization fluctuations. The term $\gamma$ is proportional to $\mu$, and 
thus is set by the OPMs or a process intrinsic to the emission mechanism. 
Its frequency dependence may be determined by the spectral index of the 
individual modes (e.g. Karastergiou et al. 2005; Smits et al. 2006; Johnston 
et al. 2008) or that of the emission's intrinsic polarization. As can be seen 
from Figure~\ref{fig:psf}, the depolarization caused by pure fluctuations alone
(the Fisher distribution from Case 1, top right panel) forms a rough, upper 
envelope to the depolarization caused by the other fluctuation geometries. 
The depolarization becomes more substantial in Cases 2 and 3 (bottom panels 
of the Figure) as the value of $\gamma$ decreases. In other words, the 
depolarization is influenced more by the intrinsic effects represented by 
$\gamma$ than by the polarization fluctuations represented by $\kappa$. 
Therefore, the main conclusion to draw from this polarization stability 
analysis is pulsar-intrinsic effects are more effective at depolarizing 
the emission than random fluctuations in the orientation of the polarization 
vector.

\section{POSSIBLE ORIGIN OF PERPENDICULAR FLUCTUATIONS}
\label{sec:perp}

A single, physical process is not likely to be responsible for all the
polarization patterns we observe. As mentioned in the Introduction, the
origin of the simple polarization patterns is qualitatively understood, 
particularly in the case of OPMs where the patterns are formed by 
fluctuations parallel to the mode polarization vectors. The origin of 
the more complicated patterns created by perpendicular polarization 
fluctuations is not understood. But as discussed below, they may arise 
from a stochastic version of GFR or scattering in the pulsar magnetosphere.

\subsection{Stochastic Version of Generalized Faraday Rotation}

In general terms, Faraday rotation is the physical process that alters the 
difference between the phases of the modes as they propagate through a plasma 
(Melrose 1979), regardless of the particle energy in the plasma, the strength 
of the magnetic field threading the plasma, or the coherence of the modes. 
The modes are incoherent when the difference in their phases at a given 
wavelength is large ($\Delta\chi\gg 1$) and are coherent (coupled) as long 
as the phase difference is small ($\Delta\chi < 1$). The modes retain their 
individual polarization identity in an observation when they are incoherent, 
but effectively lose their identity when they are coherent. Faraday rotation 
can become stochastic when the fluctuations in phase difference are large 
($\sigma_\chi \gg 1$), in which case $\Delta\chi$ can be treated as a random 
variable (Lee \& Jokipii 1975; Simonetti et al. 1984; Melrose \& Macquart 1998). 

GFR alters the component of the radiation's polarization vector that is 
perpendicular to the polarization vectors of the plasma's wave propagation 
modes. For any plasma, the unit vectors representing the polarization states 
of the two modes are anti-parallel on a diagonal through the Poincar\'e 
sphere. For the cold, weakly-magnetized plasma that is the interstellar 
medium (ISM), the propagation modes are circularly polarized, and the mode 
diagonal defined by their polarization vectors connects the poles of the 
Poincar\'e sphere. Faraday rotation in the ISM causes the orientation of 
the radiation's polarization to vary in a plane perpendicular to the mode 
diagonal, either on the Poincar\'e sphere's equator or on a small circle 
parallel to it, depending upon the polarization state of the plasma-incident 
radiation. For the relativistic plasma in the strong magnetic field of a 
pulsar's magnetosphere, the modes are thought to be linearly polarized 
(Melrose 1979; Allen \& Melrose 1982; Barnard \& Arons 1986) so that the 
mode diagonal lies in the equator of the Poincar\'e sphere. The modes in 
a relativistic, magnetized plasma of a synchrotron radio source or pulsar 
wind are, in general, elliptically polarized (Kennett \& Melrose 1998; 
Pacholczyk \& Swihart 1970; Sincell \& Krolik 1992). In these latter cases, 
GFR causes the polarization vector to rotate on a small circle in the 
Poincar\'e sphere that is perpendicular to and centered on the mode 
diagonal (e.g.  Figure 3 of Kennett \& Melrose 1998). If the modes are 
coherent, GFR causes the amplitude of the linear and circular polarization 
to vary periodically (Pacholczyk \& Swihart 1970; Cocke \& Pacholczyk 1976). 

For GFR to occur, the polarization of the radiation incident on the plasma 
must be different from the polarization of the plasma's wave propagation 
modes (e.g. Pacholczyk \& Swihart 1970). How this might occur in a pulsar's 
magnetosphere can be understood if we view the magnetosphere as being 
composed of layers, where the properties of the wave propagation modes are 
constant within a given layer, but different between layers. The mode 
properties, such as their indices of refraction and polarization states, 
are likely to vary quickly with distance, $r$, from the center of the pulsar, 
because both the particle density and magnetic field strength are thought to 
decrease as $r^{-3}$ (Goldreich \& Julian 1969). For simplicity, let us 
assume the radiation is generated in the lowest magnetospheric layer and its 
polarization state is identical to that of the wave modes in the layer to 
ensure efficient coupling between the generation and propagation of the 
radiation. The radiation's polarization is unaffected in the lowest layer, 
but gets slightly altered by GFR in subsequent layers, because the 
polarization of the radiation incident on each layer is different from the 
polarization of that layer's propagation modes. The radiation's polarization 
in each layer then contains a component that is perpendicular to the mode 
polarization because of GFR. Ultimately, a polarization limiting region 
(PLR) is reached in the upper layers of the magnetosphere where GFR is 
no longer effective in altering the radiation's polarization (Melrose 1979; 
Kennett \& Melrose 1998). The PLR occurs where, or when, $\Delta\chi = 1$. 
At or just beyond the PLR, the radiation must couple to the ISM wave modes 
for propagation through the ISM (Stinebring 1982). With this simple cartoon, 
we see how GFR may be capable of producing a polarization component that is 
perpendicular to the polarization of the wave propagation modes.

The relatively long wavelength ($\lambda > 10$ cm) observations of individual 
pulse polarization conducted to date suggest that the wave propagation 
modes in pulsar radio emission are incoherent (MS1; MS2). Consequently, the 
differences in mode phases must be large, implying GFR is operative in the 
pulsar magnetosphere. Furthermore, the emission's intensity and polarization 
are highly variable, most likely due to rapid changes, both spatially and 
temporally, in the flowrate and physical properties of the magnetospheric 
plasma. This in turn suggests that the fluctuations in the mode phase 
difference, $\sigma_\psi$, are also very large. We are then led to the 
prospect that GFR in the pulsar magnetosphere is likely to be stochastic. 
This interpretation is consistent with the analysis in \S\ref{sec:patterns} 
where the orientation angles of the polarization vectors are treated as 
random variables.  

The theory of stochastic Faraday rotation in the ISM is well-developed
(e.g. Spangler 1982; Simonetti et al. 1984; Melrose \& Macquart 1998).
The theory probes a two dimensional problem, in the Stokes parameters
Q and U, where the fluctuations in the position angle of the linear
polarization vector map directly to the fluctuations in the mode phase
difference. The extension of the theory to the more general, three 
dimensional case (Q, U, and V) of stochastic GFR has not been made and 
is beyond the scope of this paper. In the absence of the theory, knowing 
that the conditional densities derived in \S\ref{sec:patterns} are 
reasonable approximations to the polarization patterns we observe, and 
realizing that stochastic GFR may occur in the magnetosphere, we are faced 
with the possibility that some subset of these conditional densities, or 
some other joint probability density, may describe how GFR causes the 
orientation of a pulsar's polarization vector to fluctuate. Interestingly, 
the polarization stability analysis in \S\ref{sec:depol} produces the same 
conclusions developed by Melrose \& Macquart (1998) in their assessment of 
depolarization by stochastic Faraday rotation in the ISM. Specifically, 
when the conditional densities are used to calculate the moments of the 
Stokes parameters, the first moments decay and the sum of their second 
moments remains constant. In fact, the radiative transfer equation for 
the Stokes parameters (e.g.  Equation 1 of Kennett \& Melrose 1998) requires 
the second moment of the polarization, ${\rm p^2 = q^2 + u^2 + v^2}$, to 
remain constant as the radiation propagates through the plasma. All of 
the conditional and joint probability densities listed in 
\S\ref{sec:patterns} and the Appendix comply with this requirement.

\subsection{Scattering in the Magnetosphere}

Edwards \& Stappers (2004) simulated the polarization patterns in the core 
component of PSR B0329+54. They suggested that the modes are not precisely 
orthogonal, and retained the assumption of superposed modes. More importantly,
and in contrast to the models discussed in \S\ref{sec:patterns}, they also
proposed that the orientation angles for the polarization vector of one 
mode are highly dispersed while the angles of the other mode are not. They 
attributed the orientation angle fluctuations to GFR. More specifically, 
however, their model is consistent with any mechanism that selectively alters 
the orientation angles of only one mode. Melrose et al. (2006) also modeled 
the polarization patterns in the core component of PSR B0329+54 using 
assumptions similar to those of Edwards \& Stappers. They invoked the 
non-orthogonality of the modes and required the fluctuations in the 
orientation angles of one mode to be different from those in the other.
But unlike Edwards \& Stappers, they suggested that the modes are disjoint 
(i.e. they occur separately, not simultaneously) part of the time, and did 
not specifically advocate a physical mechanism for the cause of the orientation 
angle fluctuations.

One mechanism that may alter the orientation angles of one polarization mode 
and not the other is scattering in the pulsar magnetosphere (Blandford \& 
Scharlemann 1976; Sincell \& Krolik 1992; Lyubarskii \& Petrova 1996; 
Petrova 2008). Induced scattering may occur in the magnetosphere because 
large photon occupation numbers are implied by the high brightness 
temperatures observed in the radio emission. When the radiation frequency, 
$\omega$, is much less than the electron gyrofrequency, $\omega_B$, as might 
be expected near the stellar surface, the only wave propagation mode that has 
a non-zero scattering cross section is the ordinary mode (O-mode), which is 
polarized in the plane defined by the wave vector, k, and the ambient magnetic 
field, B (Blandford \& Scharlemann 1976; Sincell \& Krolik 1992). This occurs 
for two reasons. First, the extreme strength of the field ($B\simeq 10^{12}$ G) 
causes the electrons in the plasma to occupy their lowest Landau level, 
such that their motion is constrained along the field, like a bead on a 
wire. Second, in an overly simplistic description, only an incident O-mode 
wave can accelerate an electron along the field, thus causing it to radiate,
because it is the only mode having a component to its electric field that 
is parallel to the ambient magnetic field. The extra-ordinary mode (X-mode) 
cannot accelerate an electron along the field because its electric field is 
always perpendicular to the magnetic field. The scattering cross section 
varies as $\sin^2\theta$ (Blandford \$ Scharlemann 1976), where $\theta$ is 
now the angle between k and B, and no scattering will occur if k and B are 
parallel ($\theta=0$). The observed polarization of the scattered radiation 
depends upon the geometry of the magnetic field in the scattering region, 
as projected on the plane of the sky, and the temporally and spatially varying 
distribution of charged particles along the magnetic field lines.

Petrova (2008) has suggested that induced scattering at different altitudes
within the magnetosphere may lead to a depolarization of the radiation. The 
scattered radiation from different altitudes may have different position
angles, and it is the superposition of these waves with different position 
angles that leads to the depolarization. This scenario is qualitatively
consistent with what is described in \S\ref{sec:patterns}, Cases 3 and 4.

In summary, the empirical models appear to place three requirements on the 
physical processes that create the more complex polarization patterns we 
observe. First, the polarization must have a large modulation index, through a 
combination of large polarization fluctuations and a small mean polarization.
Second, the processes must create fluctuations in polarization that are
both parallel and perpendicular to the mean polarization vector. Third,
in some cases, the process may create fluctuations in the orientation of 
the polarization vector of one orthogonal mode that exceed those in the
other mode.

\section{DISCUSSION}
\label{sec:discuss}

A fundamental question remains for the observation of PSR B0329+54 by Edwards 
\& Stappers (2004). Why does the polarization annulus appear in the core 
component of the pulsar but not in its conal outriders? The annulus must be 
intrinsic to the pulsar because a propagation effect in the pulsar wind or 
the ISM presumably would affect the core and cone emission in a similar 
fashion. Explanations for the difference may reside in the distinction 
between core and cone emission made by Rankin (1983; 1990). Cone emission 
has a moderate spectral index and can be highly linearly polarized. It is 
thought to originate high above the pulsar polar cap. Core emission, on the 
other hand, has a steep spectral index, and its polarization signature is 
often a change in the handedness of the circular polarization near the core 
peak. It is thought to originate at or near the polar cap. The core and cone 
radiation from PSR B0329+54 appear to follow this general characterization 
of the emission components (Karastergiou et al. 2001). We can surmise that 
GFR occurs throughout the pulse of PSR B0329+54 because observations (Edwards 
\& Stappers 2004) indicate that independent OPMs occur at most locations within 
its pulse. Of course, the core emission would be more susceptible to GFR 
because its propagation path length is presumably much larger than that for 
the cone emission. Stochastic GFR may occur predominantly near the stellar 
surface where a turbulent plasma outflow causes fluctuations in the mode 
phase difference. Stochastic GFR may get suppressed in the propagation region 
for the cone emission because the cross section of the plasma flux tube 
increases with distance from the star and, thus, the plasma outflow becomes 
more laminar. Alternatively, a popular model for pulsar radio emission (e.g. 
Ruderman \& Sutherland 1975) calls for an intense photon beam to be created 
by primary charged particles that are accelerated in a voltage potential gap 
near the stellar surface. These photons produce secondary pairs of charged 
particles that go on to radiate via coherent curvature radiation, for example. 
In this scenario, perhaps the polarization annulus appears only in the core 
emission because only the photon beam can provide the large photon occupation 
numbers that are conducive to induced scattering. Scattering would be most 
evident where the wave vector of the O-mode radiation is perpendicular to 
the ambient magnetic field, as might be expected in the multi-polar structure 
of the field near the stellar surface where the core emission is thought to 
orginate. Furthermore, the specific scattering process described in 
\S\ref{sec:perp} will not occur higher in the magnetosphere where the 
electrons can occupy higher Landau levels and the cyclotron frequency can 
approach the radiation frequency. Finally, the presence of both the 
polarization annulus and the sign-changing circular polarization in the core 
component of PSR B0329+54 begs a secondary question of whether the two 
phenomena are related or entirely coincidental. Single pulse observations 
of other pulsars dominated by core emission, such as PSR B1933+16, could 
test whether the phenomena are associated with one another.

The analytical model of pulsar polarization presented in this paper can 
describe most, but not all, of the polarization patterns observed in PSR 
B0329+54. In the cone emission of the pulsar, the LEAPs show a data cluster 
with circular cross-section in each hemisphere, consistent with the pattern 
described in Case 2. The cluster in the precursor to the core component has 
the shape of an elliptical bar, which is the pattern produced by Case 4. 
The LEAP observed at the core component shows an elliptical bar in one 
hemisphere and a partial annulus in the other. None of the cases considered 
in \S\ref{sec:patterns} can reproduce this pattern. Additional assumptions, 
such as those described by Melrose et al. (2006) and Edwards \& Stappers
(2004), must be invoked to model the polarization at this pulse location.
 
The oft-stated objective of polarization observations of individual pulses 
(e.g. Manchester et al. 1975; Stinebring et al. 1984) is to understand the 
radio emission mechanism of pulsars. Progress has been made on this front, 
particularly with total intensity measurements that reveal a carousel of 
subbeams circulating about the star's magnetic pole (e.g. Deshpande \& 
Rankin 1999). But in most cases, propagation effects in the pulsar 
magnetosphere, not radio emission mechanisms, are invoked to interpret the 
results of these observations. Afterall, a multitude of propagation effects 
in the ISM complicates our reception of the pulsar signal, so we should not 
be surprised that propagation effects in the pulsar magnetosphere complicate 
our view of what happens there. For example, the occurrence of OPMs has been 
attributed to the birefringence of the magnetospheric plasma above the pulsar 
polar cap (Allen \& Melrose 1982; Barnard \& Arons 1986; Petrova 2001). 
Cyclotron absorption high in the magnetosphere (Luo \& Melrose 2001; 
Melrose 2003) has been proposed as a possible origin of circular polarization 
in the emission. In this paper, GFR and scattering are suggested as possible 
candidates for the origin of some polarization patterns. Individually and 
collectively, these propagation effects can obscure the polarization of the 
underlying emission mechanism. If these interpretations are correct, they 
imply that single pulse polarization observations are best suited for probing 
propagation effects in pulsar magnetospheres. The emission mechanism may 
reveal its secrets through total intensity observations, similar to the high 
time resolution measurements by Hankins \& Eilek (2007) who recently found 
multiple, narrow, radiation bands in the emission from the Crab pulsar.

\section{CONCLUSIONS}
\label{sec:conclude}

An empirical, analytical model of pulsar polarization has been generalized
to accommodate a wide variety of polarization fluctuation geometries. The 
model is based upon the proposition that the observed polarization of 
pulsar radio emission is due to the incoherent superposition of highly 
polarized orthogonal modes. For the modes to propagate independently,
generalized Faraday rotation may be operative in the pulsar magnetosphere 
for the modes to get significantly out of phase. The model replicates the 
polarization patterns observed in many objects and reproduces the numerical 
results from other work. When the fluctuations are parallel to the 
polarization vectors of the wave propagation modes, the patterns consist 
of two tight clusters, each in a separate hemisphere of the Poincar\'e 
sphere. The patterns assume shapes of bars, bow ties, and annuli when the 
fluctuations are perpendicular to the vectors. The more interesting patterns 
occur when the polarization modulation index exceeds unity. The diverse 
polarization patterns are not likely to originate from the same physical 
process. The parallel polarization fluctuations are caused by fluctuations 
in the polarized intensities of the orthogonal modes. The perpendicular 
fluctuations may be caused by a stochastic version of generalized Faraday 
rotation, which would require large fluctuations in the difference between 
mode phases. An expansion of the two dimensional theory of stochastic 
Faraday rotation in the ISM to the three dimensional case for pulsar 
magnetospheres may aid the interpretation of the observed polarization 
patterns. An alternative model suggests that one mode may experience 
fluctuations perpendicular to its polarization vector while the other 
does not, implying the presence of a mode-selective, random process, such 
as scattering in the pulsar magnetosphere. The polarization patterns reflect 
polarization instabilities that can depolarize the emission in a way that 
is similar to stochastic Faraday rotation in the ISM. The depolarization 
has been quantified with a polarization stability factor for the simpler 
fluctuation geometries. The stability factors imply that pulsar-intrinsic 
effects are more effective in depolarizing the emission than fluctuations 
in the orientation of its polarization vector. For all geometries evaluated 
with the model, the joint probability density of the polarization vector's 
orientation angles follows the same functional form apart from parameters 
determined by the geometry of the polarization fluctuations. The conditional 
density of the orientation angles in all cases follows the Fisher and 
Bingham-Mardia family of distributions. 

\acknowledgements

I thank Dan Stinebring for providing the data used in the analysis. 

\clearpage

\begin{center}
\bigskip
{\bf APPENDIX\\
\medskip
Joint Probability Density of Colatitude and Longitude}
\end{center}

The basic functional form of the joint probability density of a polarization 
vector's colatitude, $\theta$, and longitude, $\phi$, for all cases considered 
in \S\ref{sec:patterns} is given by

\begin{equation}
g(\theta,\phi) = z{\rm C}{\sin\theta\over{4\pi}}
 \Biggl\{\exp{\Biggl({y^2\over{2}}\Biggr)} \Biggl[1+{\rm erf}
 \Biggl({y\over{\sqrt{2}}}\Biggr)\Biggr](1 + y^2)+y\sqrt{{2\over{\pi}}}\Biggr\},
\label{eqn:joint}
\end{equation}

\noindent where $\rm{erf}(x)$ is the error function, C is a constant, and the
parameters $y$ and $z$ are functions of $\theta$ and $\phi$ determined by the 
geometry of the polarization fluctuations. The analytical expressions for C, 
$y$, and $z$ for each case are listed below. 

{\bf Case 1:}

\begin{equation}
{\rm C} = \exp\Biggl(-{s^2\over{2}}\Biggr)
\end{equation}

\begin{equation}
y(\theta,\phi) = s\cos\theta
\end{equation}
 
\begin{equation}
z(\theta,\phi) = 1
\end{equation}

{\bf Case 2:}
 
\begin{equation}
 {\rm C}= \exp{\Biggl[-{s^2\over{2(1+\rho^2)}}\Biggr]}
\end{equation}

\begin{equation}
y(\theta,\phi) = {s\cos\theta\over{[(1 + \rho^2\sin^2\theta)
                 (1 + \rho^2)]^{1/2}}}
\end{equation}
 
\begin{equation}
z(\theta,\phi) = {(1+\rho^2)\over{(1 +\rho^2\sin^2\theta)^{3/2}}}
\end{equation}

{\bf Case 3:}
 
\begin{equation}
{\rm C} = \exp\Biggl(-{s^2\over{2}}\Biggr)
\end{equation}

\begin{equation}
y(\theta,\phi) = s\cos\theta\Biggr({1+\eta^2
                 \over{1 + \eta^2\cos^2\theta}}\Biggl)^{1/2}
\end{equation}

\begin{equation}
 z(\theta,\phi) = {(1+\eta^2)^{1/2}\over{(1 +\eta^2\cos^2\theta)^{3/2}}}
\end{equation}

{\bf Case 4:}
 
\begin{equation}
{\rm C} = \exp\Biggl({-s^2\over{2}}\Biggr) 
\end{equation}

\begin{equation}
y(\theta,\phi) = s\cos\theta\Biggr[{1 + \eta^2\over{1 + \eta^2
   (\cos^2\theta +\sin^2\theta\sin^2\phi)}}\Biggl]^{1/2}
\end{equation}
 
\begin{equation}
z(\theta,\phi) = {(1+\eta^2)
                 \over{[1+\eta^2(\cos^2\theta+\sin^2\theta\sin^2\phi)]^{3/2}}}
\end{equation}

{\bf Case 5:}
 
\begin{equation}
{\rm C} = \exp\Biggl[{-s^2\over{2(1+\rho^2)}}\Biggr] 
\end{equation}

\begin{equation}
y(\theta,\phi) = s\cos\theta
                  \Biggr({1+\eta^2\over{1+\rho^2}}\Biggr)^{1/2}
                  \Biggr\{{1\over{(1+\eta^2) + \sin^2\theta
                 [(\rho^2-\eta^2)+\eta^2(1+\rho^2)\sin^2\phi]}}\Biggl\}^{1/2}
\end{equation}

\begin{equation}
z(\theta,\phi) = {(1+\eta^2)(1+\rho^2)\over{\{(1+\eta^2)
       + \sin^2\theta[(\rho^2-\eta^2)+\eta^2(1+\rho^2)\sin^2\phi]\}^{3/2}}}
\end{equation}

\noindent Case 5 provides a general joint probability density for $\theta$ 
and $\phi$. It becomes the joint density for Case 1 when $\eta=\rho=0$, for 
Case 2 when $\eta=0$, and for Case 4 when $\rho=0$. When $s=0$, the joint 
probability density becomes $g(\theta,\phi)=z(\theta,\phi)\sin\theta/4\pi$ 
for all cases since both C and the bracketed term in Equation~\ref{eqn:joint} 
become equal to one.

\begin{deluxetable}{ccccc}
\tablenum{1}
\tablewidth{450pt}
\tablecaption{Summary of Fluctuation Geometry Cases}
\tablehead{
 \colhead{Case} & \colhead{Fluctuations} & \colhead{Cluster Shape} & 
  \colhead{Axis Orientation} & \colhead{Conditional Density}}
\startdata
 1 & $\sigma_q=\sigma_u=\sigma_v$ & Spheroid & NA & Fisher  \\
 2 & $\sigma_q=\sigma_u<\sigma_v$ & Prolate Ellipsoid& Parallel & BM bipolar \\
 3 & $\sigma_q=\sigma_u>\sigma_v$ & Oblate Ellipsoid& Parallel & BM girdle \\
 4 & $\sigma_q>\sigma_u=\sigma_v$ & Prolate Ellipsoid& Perpendicular & 
      BM hybrid\\
 5 & $\sigma_u<\sigma_q,\sigma_v$ & Irregular Ellipsoid & $\sigma$-dependent
   & BM hybrid\\
\enddata
\end{deluxetable}

\end{document}